\documentclass[aps,prx,reprint,superscriptaddress,showpacs,showkeys, longbibliography]{revtex4-1}

\usepackage{latexsym,amssymb,amsfonts,mathrsfs,color,graphicx,amsmath}
\usepackage[usenames,dvipsnames]{xcolor}
\usepackage{soul}
\usepackage[linktocpage,colorlinks=true,linkcolor=blue,citecolor=blue,breaklinks=true,urlcolor=blue]{hyperref}

\usepackage[separate-uncertainty = true,multi-part-units=single]{siunitx}

\usepackage{textcomp}

\usepackage{sidecap}

\newcommand{\derpar}[2] {\frac{\partial #1}{\partial #2}}
\newcommand{\reff}[1] {(\ref{#1})}

\begin{document}

\title{3D spatial exploration by \textit{E. coli} echoes motor temporal variability}

\author{Nuris Figueroa-Morales}
\affiliation{PMMH, UMR 7636 CNRS-ESPCI-PSL Research University, Sorbonne University,
University Paris Diderot, 7-9 quai Saint-Bernard, 75005 Paris, France}
\affiliation{Department of Biomedical Engineering, The Pennsylvania State University, University Park, PA 16802, USA}

\author{Rodrigo Soto}
\affiliation{Departamento de F\'\i sica, FCFM, Universidad de Chile, Santiago, Chile}

\author{Gaspard Junot}
\affiliation{PMMH, UMR 7636 CNRS-ESPCI-PSL Research University, Sorbonne University,
University Paris Diderot, 7-9 quai Saint-Bernard, 75005 Paris, France}

\author{Thierry Darnige}
\affiliation{PMMH, UMR 7636 CNRS-ESPCI-PSL Research University, Sorbonne University,
University Paris Diderot, 7-9 quai Saint-Bernard, 75005 Paris, France}

\author{Carine Douarche}
\affiliation{Laboratoire de Physique des Solides, CNRS, Universit\'e Paris-Sud, Universit\'e Paris-Saclay, 91405 Orsay Cedex, France}

\author{Vincent Martinez}
\affiliation{Institute of Condensed Matter and Complex System, The University of Edinburgh, Edinburgh, EH9 3FD, UK}

\author{Anke Lindner}
\affiliation{PMMH, UMR 7636 CNRS-ESPCI-PSL Research University, Sorbonne University,
University Paris Diderot, 7-9 quai Saint-Bernard, 75005 Paris, France}

\author{\'Eric Cl\'ement}
\email{eric.clement@upmc.fr}
\affiliation{PMMH, UMR 7636 CNRS-ESPCI-PSL Research University, Sorbonne University,
University Paris Diderot, 7-9 quai Saint-Bernard, 75005 Paris, France}


\begin{abstract}
Unraveling bacterial strategies for spatial exploration is crucial for understanding the complexity in the organization of life. Bacterial motility determines the spatio-temporal structure of microbial communities, controls infection spreading and the microbiota organization in guts or in soils.
Most theoretical approaches for modeling bacterial transport rely on their run-and-tumble motion.
For Escherichia coli, the run time distribution was reported to follow a Poisson process with a single characteristic time related to the rotational switching of the flagellar motors. 
However, direct measurements on flagellar motors show heavy-tailed distributions of rotation times stemming from the intrinsic noise in the chemotactic mechanism. 
Currently, there is no direct experimental evidence that the stochasticity in the chemotactic machinery affect the macroscopic motility of bacteria.
In stark contrast with the accepted vision of run-and-tumble, here we report a large behavioral variability of wild-type \emph{E. coli}, revealed in their three-dimensional trajectories.
At short observation times, a large distribution of run times is measured on a population and attributed to the slow fluctuations of a signaling protein triggering the flagellar motor reversal.
Over long times, individual bacteria undergo significant changes in motility.
We demonstrate that such a large distribution of run times introduces measurement biases in most practical situations.
Our results reconcile the notorious conundrum between run time observations and motor switching statistics. We finally propose that statistical modeling of transport properties currently undertaken in the emerging framework of active matter studies, should be reconsidered under the scope of this large variability of motility features.

\end{abstract}

\maketitle

\section{Introduction}
 
The \emph{run-and-tumble} (R\&T) strategy developed by bacteria for exploring their environment is a cornerstone of quantitative modeling of bacterial transport.  
In this paradigm, bacteria swim straight during a \emph{run-time}, undergo a reorientation process during a \emph{tumbling-time} and pursue thereafter the next run in a different direction. The now standard vision of the R\&T strategy was established in the 70's for swimming \textit{E. coli} by Berg and Brown \cite{Berg1972, Berg2004}, based on 3D trajectories obtained via a Lagrangian tracking technique. They proposed that an adapted bacterium would perform, over long times, an isotropic random walk composed of the run and tumble phases, both distributed in time as a Poisson process \cite{Berg2004, Berg1972, Saragosti2011, alon1998response, berg1993random}. 
For quantitative analysis, the run-time and tumble-time distributions are often taken as Poisson processes with typical values $\overline{\tau}_{run} \sim \SI{1}{\second}$ and $\overline{\tau}_{tumble} \sim \SI{1/10}{\second}$  \cite{Berg2004,Qu2018}. These values change in the presence of chemical gradients, leading to a biased random walk known as chemotaxis.

Alongside the relevance of this result in the context of biology, medicine or ecology, fluids laden with motile bacteria have become an epitome for active matter, where the organization of active particles recently led scientists to revisit many concepts of out-of-equilibrium statistical physics \cite{wu2000particle, saintillan2013active, ReviewActiveMatter2013, Solon2015}. Suspensions of motile bacteria are systems of choice for these studies \cite{schwarz-linek2016escherichia} and many original phenomena such as anti-Fick's law migration \cite{galajda2007wall}, collective motion \cite{dombrowski2004self}, viscosity reduction \cite{Sokolov2009viscosity, Gachelin2014, Matias2015}, enhanced diffusion \cite{wu2000particle} or motion rectification \cite{Sokolov2010, DiLeonardo2010, Wioland2013, Kaiser2014} have been discovered.
Most recent theoretical studies on active matter, aimed at understanding the emergence of collective motion or other macroscopic transport processes in bacterial fluids, assume uncorrelated orientational noise, which is the direct consequence of the Poisson character of the R\&T process \cite{ReviewActiveMatter2013}.

\begin{SCfigure*}
    \includegraphics[width=1.5\linewidth]{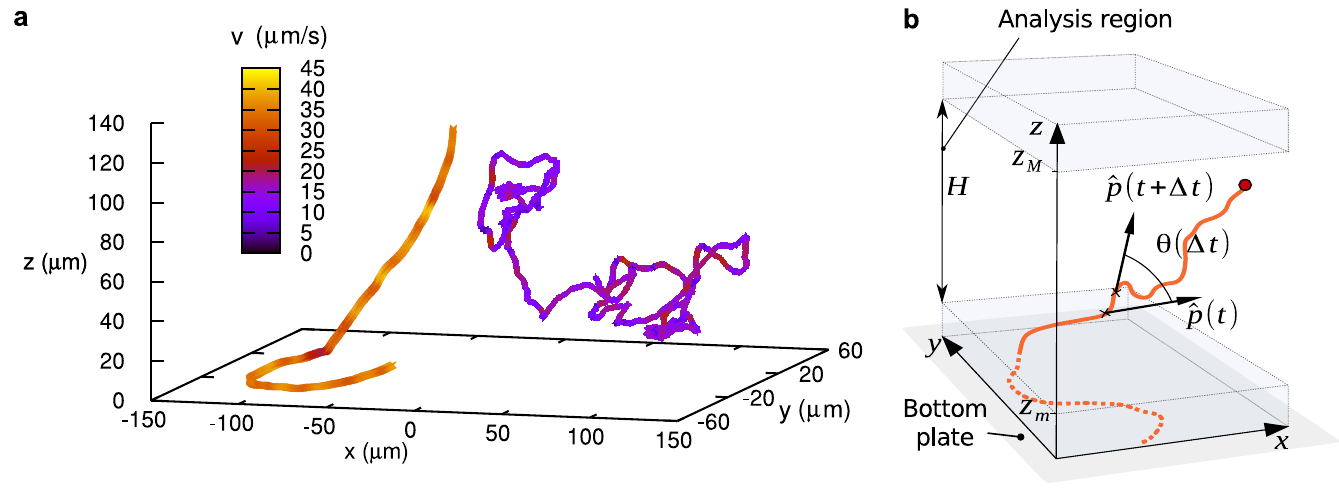}
    \caption{Lagrangian 3D tracking of bacteria and analysis conditions. (a) RP437 wild-type \textit{E.\,coli} displaying very different typical trajectories: persistent trajectory (Bact 1: $\tau_p= \SI{12}{\second}$) and non-persistent trajectory (Bact 2: $\tau_p=\SI{0.7}{\second}$). (b) Sketch of the part of the track used for analysis and angles used for computing $C(\Delta t) =  \langle \hat{p}(t)\cdot\hat{p}(t+\Delta t)\rangle = \langle \cos \theta(\Delta t)\rangle $, using a sliding window for average on time $t$.}
    \label{fig:two_tracks_sketch}
\end{SCfigure*}

The simple approach of introducing a Poisson distribution for the run-times, although useful for simple qualitative interpretations, is not fully consistent with a growing number of measurements performed on the individual rotary motors \cite{Korobkova2004, Korobkova2006PRL, Emonet_Cluzel2008, min2009high, wang2017non} driving the helix-shaped flagella. For \textit{E. coli}, the forward (run) motion is associated with the counterclockwise rotation (CCW) of the motors and the tumbles take place when the motors rotate clockwise (CW). The CCW to CW transition is regulated by an internal biochemical process associated with the phosphorylation of the CheY protein.

In a seminal work, Korobkova \textit{et al.}\ \cite{Korobkova2004} brought evidence for a heavy tail distribution for the duration of CCW rotations. Importantly, this highlights possible coupling between the stochastic fluctuations in the chemotactic biochemical network and the emergent bacterial motility. 
Consequences could affect the macroscopic organization of bacterial populations, chemotactic response to chemical heterogeneity and also, genetic and epigenetic feedback of bacterial populations to environmental constraints.

Its potential importance in the context of active matter studies remains overlooked. For multi-flagellated bacteria, the correspondence between switching statistics, motor synchronization, flagellar bundling/unbundling dynamics and finally, large-scale exploration properties, remains unclear. 
Currently, there is no direct experimental evidence
that the macroscopic motility of free swimming bacteria is sensitive to the stochasticity borne by the chemotactic biological circuit. This is precisely the question we address here. 

Conceptually, our analysis starts from 
the extreme sensitivity of the rotational CCW$\rightarrow$CW switching to the abundance of the phosphorylated protein CheY-P in the cell. 
This picture induces a time-scale separation since, at short times, the alternation of CCW and CW rotations keeps memory of a quasi-fixed level of CheY-P. This memory is erased at longer times and we thus expect  very different run times and motility features at the macroscopic level. 

For the first time, we link the individual motor rotation statistics to the global motility features that we observed in a large number of 3D trajectories of wild type \textit{E. coli} bacteria. 
At short observation times, the time persistence of the swimming orientations displays an exponential decay as classically admitted, but with a large distribution of characteristic times within a population of monoclonal bacteria. However, when tracking the cells individually over several tenths of minutes, we identify for each cell a large behavioral variability. 
The motility data are quantitatively analyzed  through a simple chemotactic model initially proposed by Tu and Grinstein \cite{Tu2005} involving the fluctuations of CheY-P triggering the tumbling events. The model is here adapted to render the spatial exploration process. It now explains the occurrence of a large behavioral variability of swimming direction and also why at short observation times, a large distribution of these is expected over a population. The central outcome of this model is that the persistence time durations naturally follow a log-normal distribution, instead of a standard Poisson distribution. 
Importantly, we identify a source of measurement bias introduced in most practical situations, that is a consequence of such a large distribution of run-times. 
Finally, we discuss the consequences of measuring averaged quantities over a population displaying a large distribution of motility features. This source of measurement bias is relevant in the general framework of experiments on statistical physics of active matter.

\section{Variability of bacterial motility in a population} \label{sec.population}

To characterize the bacterial motility, we  built an automated tracking device suited to follow fluorescent objects and record their 3D trajectories. A swimming bacterium is kept automatically in the center of the visualization field and at the focus of an inverted microscope by a visualization feed-back loop acting horizontally on a mechanical stage and vertically on a piezo stage.
The method is fully detailed in reference \cite{Darnige2016} by Darnige \textit{et al.} (see also Materials an Methods) and was  recently used to investigate the swimming of bacteria in a Poiseuille flow~\cite{junot2019swimming}. 

We first monitor more than a hundred swimming \textit{E.\,coli} from different strains (see Materials and Methods) in homogeneous diluted suspensions (concentration $\sim \SI{e5}{bact.\milli\liter}$) confined between two horizontal glass-slides, $\SI{250}{\micro\meter}$ apart. 
Fig.\ \ref{fig:two_tracks_sketch}(a) shows two typical trajectories from the same batch of monoclonal wild-type \textit{E.\,coli}.
We center our analysis on pieces of tracks exploring the bulk [Fig.\ \ref{fig:two_tracks_sketch}(b)] i.e., in a measurement region located \SI{10}{\micro\meter} above the surface and of maximum height of height $H=\SI{130}{\micro\meter}$.
For this series of experiments the duration of a track is at minimum \SI{8}{\second}.

\begin{figure*}
\centering
  \includegraphics[width=\linewidth]{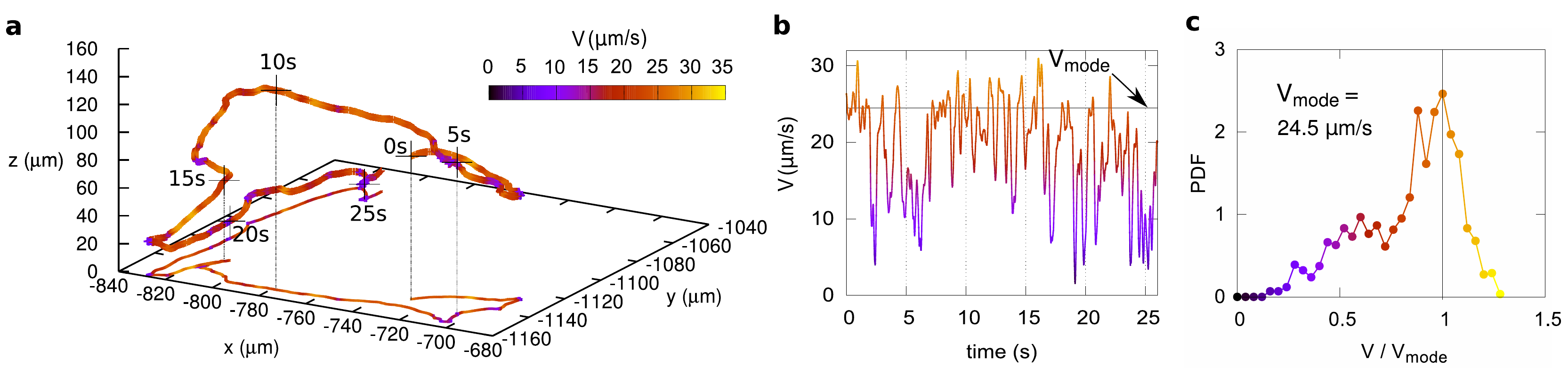}
\caption{Details of a typical trajectory. (a) 3D trajectory and its projection on the $x-y$ plane, (b) velocity vs.\ time and (c) velocity distribution. The marks every \SI{5}{\second} in the 3D track are references for comparison with panel (b).}
\label{fig:velocity_dist}
\end{figure*}

The bacterial velocities $\vec{V}(t)$ are obtained after a smoothing procedure of the trajectories over \SI{0.1}{\second}. Figure \ref{fig:velocity_dist} shows an example of a 3D trajectory and its velocity. Typically, the velocity curves for each track are irregular [Fig.\ \ref{fig:velocity_dist}(b)]. For a single track, the velocity distribution [Fig.\ \ref{fig:velocity_dist}(c)] shows a peak corresponding to the run phase and a low velocity tail that might correspond to tumbling events.
For the wild type strain RP437 in motility buffer, the average of the peak values for $V=|\vec{V}(t) |$ over the different tracks is $ \langle V \rangle=\SI{27\pm6}{\micro\meter/\second}$.

Standard analysis to extract run-time distributions relies on the identification of  tumbling events, usually done by detecting velocity drops and/or abrupt changes in swimming direction, which, without direct observation of the flagella, requires the choice of arbitrary criteria \cite{Berg2004,Qu2018}. As an illustration, Figs.\ \ref{fig:velocity_dist}(a) and (b) show that abrupt direction changes can take place without representative velocity decrease and velocity drops are sometimes not associated with reorientation.

Here, in order to characterize the motility features, we do not seek to explicitly identify the tumbling events.
We rather use the orientation correlation function $C(\Delta t)$ as a direct measurement of the swimming direction persistence. 
The director vectors pointing along the track are determined as $\hat{p}(t)= \frac{\vec{V}(t)}{V(t)}$ for each track.
For each trajectory we compute: $C(\Delta t) = \langle \hat{p}(t)\cdot\hat{p}(t+\Delta t) \rangle =\langle \cos(\theta(\Delta t)) \rangle $, where $\theta$ is the angle between swimming directors separated by a time lag $\Delta t$ [Fig.~\ref{fig:two_tracks_sketch}(b)]. The brackets denote an average over a time window sliding along the track. 
To ensure good statistics, the maximum time lag time $\Delta t$  
is chosen as one-tenth of the total track duration.
The orientation correlation reflects the R\&T statistics, but advantageously does not require an ad-hoc criterion. In Fig.~\ref{fig:correlation}(a), 30 orientation correlation functions obtained from separate tracks of different bacteria (RP437 wild-type in M9G) are displayed as a function of $\Delta t$.

From the classical picture of an exponential distribution of run times, the orientation correlation function is expected to decay exponentially with a typical decay time of $\tau_\text{p}$, defining the persistence time of the trajectory. For a characteristic run time of $\overline{\tau}_\text{run}=\SI{1}{\second}$  and a distribution of reorientation angles of mean value $\theta_\text{m} = \SI{51}{\degree}$   \cite{Berg1972} one finds $\tau_\text{p} = \frac{\overline{\tau}_\text{run}}{1-\langle\cos(\theta)\rangle}=\SI{1.5}{\second}$~\cite{Lovely1975}. Recently, a slight dependence of this angle on the swimming speed was demonstrated \cite{taute2015high}, but will be neglected in our study. 
Taking into account rotational Brownian diffusion during the run phase also leads to an exponential decaying correlation function (see Appendix A), but its contribution represents a slight modification to $\tau_\text{p}$ due to the much longer time scales of Brownian diffusion.
The predicted correlation function is represented by the dotted line on Fig.~\ref{fig:correlation}(a). Strikingly, the experimental curves display a broad scattering indicating a very large distribution of persistence times within this monoclonal population of bacteria.

Fitting the correlation functions with an exponential decay $\exp(-\tau/\tau_\text{p})$, we determine the persistence times $\tau_\text{p}$ for each track. In Fig.~\ref{fig:correlation}(b), we display them on a logarithmic vertical axis  for the strain RP437 in motility buffer (MB) and MB supplemented with Serine (MB-S). In addition, persistence times obtained in a richer medium (M9G) and for a different wild type strain AB1157 in (MB-S) are shown. 
The results prove that the distribution of orientation persistence times for wild-type bacteria is very large and, within statistical errors, they are independent of strain and chemical environment (poor or rich). 
For the very persistent tracks, the observed decorrelation remains weak over the accessible time lags. The obtained persistence times thus have a significant uncertainty, but we can be sure that their decorrelation time will be at least, bigger than the time-span of the track ($ \tau_\text{p} > \SI{8}{\second}$).
Finally, we consider the strain CR20, a smooth swimmer that tumbles only very rarely. In this case the time distribution is gathered around the average $\tau_\text{p}=\SI{25 \pm 10}{\second}$, which is close to the Brownian rotational diffusion constant $\tau_\text{p}=\tau_\text{B}=1/2D_r^\text{B}$, as expected. 
This value is however, strongly dependent on the bacterial dimensions and aspect ratio \cite{perrin_1934_I, perrin_1936_II}. A bacterium modeled as an ellipsoid of semiaxes $a = \SI{4}{\micro\meter}$ and $b = c = \SI{0.4}{\micro\meter}$ will have a persistence time $\tau_\text{p} \sim \SI{22}{\second}$, while with $a = \SI{6}{\micro\meter}$ will have a persistence time three times larger, $\tau_\text{p} \sim \SI{66}{\second}$ \cite{Figueroa_Thesis}. Therefore, the wide distribution of persistence times for CR20 could arise from the bacterial size distribution. A possible origin of this dispersion on the measurement protocol is discussed in Sect.\ \ref{sec.comparisonmodel}

\section{Variability of individual bacterial motility over time}
The large diversity of trajectories here observed over short times in bacterial populations leads to the question of its origin. The diversity could arise from a phenotype multiplicity present in the monoclonal population \cite{Smits2006, Waite_Emonet_Review2018}, where each bacterium is characterized by a mean run-time; alternatively, it could be due to temporal variability of the bacterial behavior, with mean run-times varying over the course of time.
To determine which scenario is taking place, we perform a second series of measurements, where we follow individual bacteria over very long times (up to \SI{20}{\minute}). 
In the new configuration the top and bottom of the measurement chamber are within the observation range or the 3D tracker device. We follow individual bacteria as they alternate between the surfaces and the bulk, as sketched in Fig. \ref{fig:shallow_channel} (a). For the analysis, individual tracks are cut in pieces localized entirely in the bulk (\SI{10}{\micro\meter} away from the walls). For each piece we extract the persistence time from the correlation function. Finally, for each bacterium we obtain a list of persistence times as a function of time.
If the population displayed a large distribution of fixed run-times, one would expect for each bacterium a sequence of persistence times narrowly distributed around a characteristic value, but this value would be different for different bacteria. Fig.~\ref{fig:shallow_channel}(b) carries a very different message.
For each of the tracks tested, the persistence times span a range of the same magnitude as for the whole population using shorter tracking times (see Fig.\ \ref{fig:correlation}). 

Previous studies based on 3D Eulerian tracking techniques \cite{wu2006collective, taute2015high}, i.e., on a fixed reference frame or even Lagrangian tracking technique \cite{Qu2018}  were limited to short observation times and consequently were not able to catch such slow fluctuations of the run time.
The fact that for a given bacterium the sequence of persistence times is largely distributed, confirms the importance of behavioral variability in the motility process. 
However, due to tracking time limitations imposed by the bleaching of the fluorescent signal, we were not able to test precisely to which extent the behavioral variability contains features which could vary from one bacterium to the other, stemming from inherent phenotype variations, as identified for example by Dufour \textit{et al.}\ \cite{Dufour2016}.

\section{Motility and motor rotation statistics}

\begin{figure}
\centering
  \includegraphics[width=0.95\linewidth]{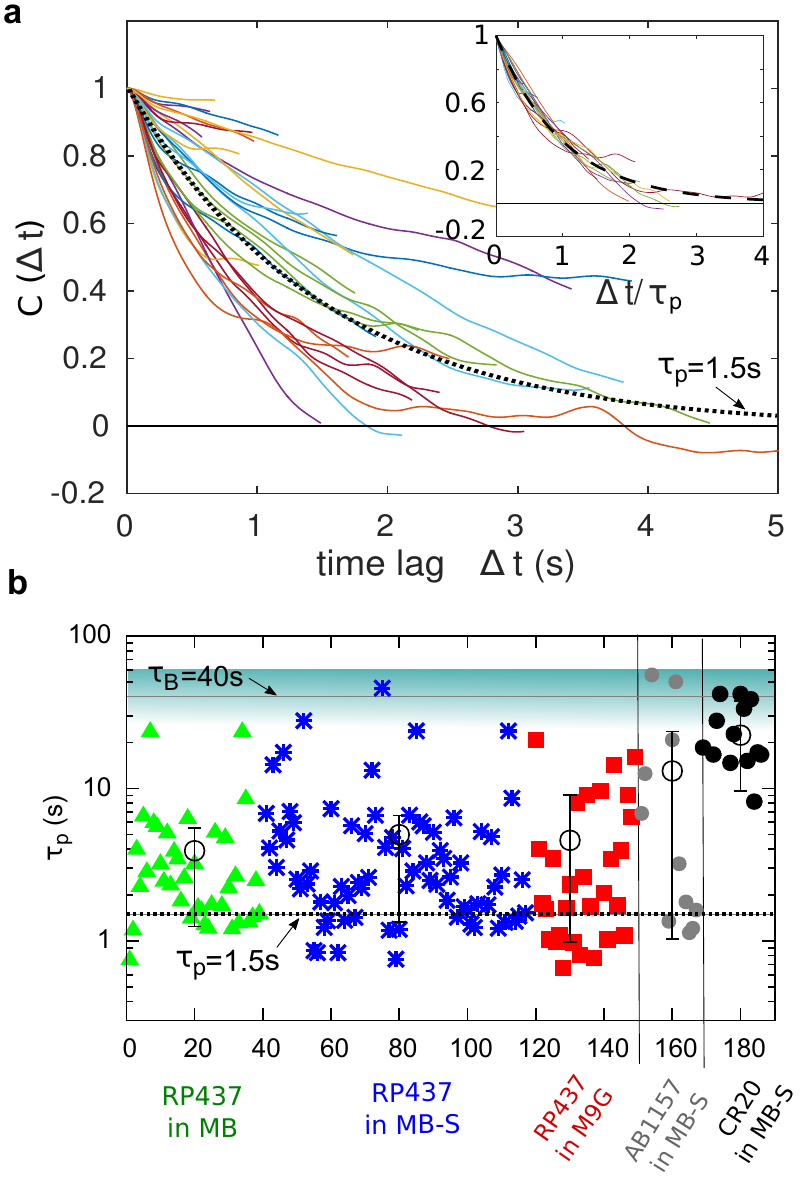}
\caption{Swimming orientation correlations. (a) Correlation function $C(\Delta t)$ obtained for 30 tracks of different RP437 bacteria in M9G, showing a large distribution of persistence times. The correlation functions are fitted with an exponential decay $\exp(-\tau/\tau_\text{p})$ to extract the persistence times $\tau_\text{p}$. The dotted line corresponds to $\tau_\text{p}$=\SI{1.5}{\second} as expected from \cite{Berg1972}. Inset: correlation functions as a function of $\Delta t$ rescaled by $\tau_\text{p}$. The dashed line is $\exp(-x)$.
(b) Persistence times for individual bacteria of wild type strains RP437 and AB1157, and smooth swimmer mutant CR20 in different media (MB, MB-S and M9G).
The blue background region designates the cutoff  from Brownian diffusion and the solid line is the average value corresponding to a bacterium of length \SI{10}{\micro\meter} ($\tau_\text{B} = \SI{40}{\second}$). The dotted line corresponds to the expected $\tau_\text{p}$=\SI{1.5}{\second} also represented in panel (a).
Uncertainty bars indicate the mean and confidence interval at $ 68\% $.
}
\label{fig:correlation}
\end{figure}

\begin{figure}
  \includegraphics[width=0.95\linewidth]{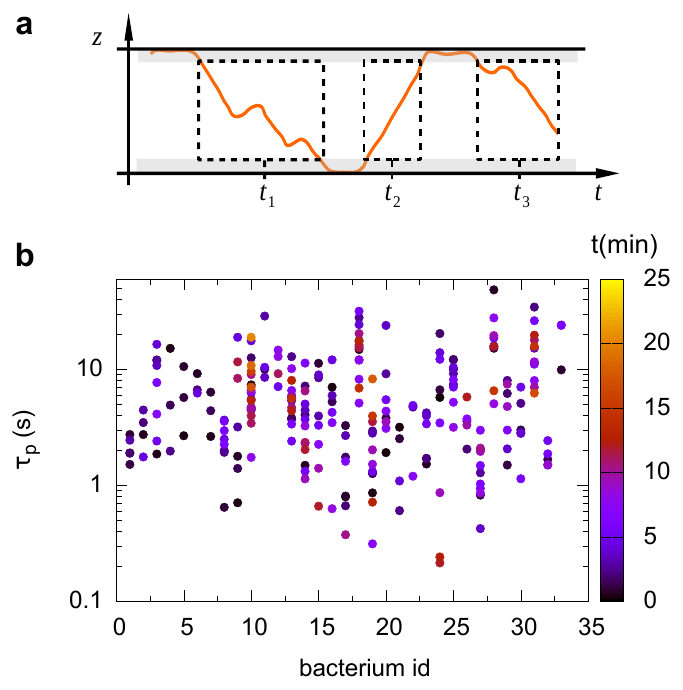}
\caption{
Analysis  for long tracks.
(a) Sketch indicating the pieces of track from the same trajectory selected for demonstrating the behavioral variability.
(b)
$\tau_\text{p}$ for pieces of track from the same trajectories, for 33 different RP437 bacteria in MB-S. The color represents the starting time of the measurement. Each bacterium displays a large variation of persistence times.
}
\label{fig:shallow_channel}
\end{figure}

The presence of a behavioral variability, as identified earlier, raises the question of its biochemical origins. Previous results point towards a definite influence of a stochastic process in the chemotactic sensory circuit. At the end of the biochemical cascade there is a phosphorilation of a CheY protein (CheY-P) promoting a switch in the motor rotation from the CCW state (run phase) to the CW state (tumbling phase). 
The most accepted picture rendering the CCW$\rightleftharpoons$CW transition is a two state model initially proposed by Khan \textit{et al.}\ \cite{Khan1980} which considers the switching of the rotation direction CCW$\rightarrow$CW (equivalently CW$\rightarrow$CCW) as an activated process regulated by the presence of CheY-P. The double well Gibbs free energy associated with the transition CCW$\rightleftharpoons$CW depends in a very sensible way on the CheY-P ($[Y]$) concentration values near the motor, as it was shown by Cluzel \textit{et al.}\ \cite{Cluzel2000}. This strong sensitivity leads naturally to behavioral variations, as slow fluctuations around the mean value can change the motility features from preferentially tumbling (high CheY-P) to preferentially running (low CheY-P). It also means that at short times the CheY-P level does not change significantly and motility features remain constant. Therefore, at a given moment the motility features should be largely distributed in a population of bacteria bearing different CheY-P concentrations. This is in essence what is observed in our experiments in Figs.\ \ref{fig:correlation} and \ref{fig:shallow_channel}.

\subsection{Quantitative description of the behavioral variability model}
\begin{figure*}
\centering
 \includegraphics[width=\linewidth]{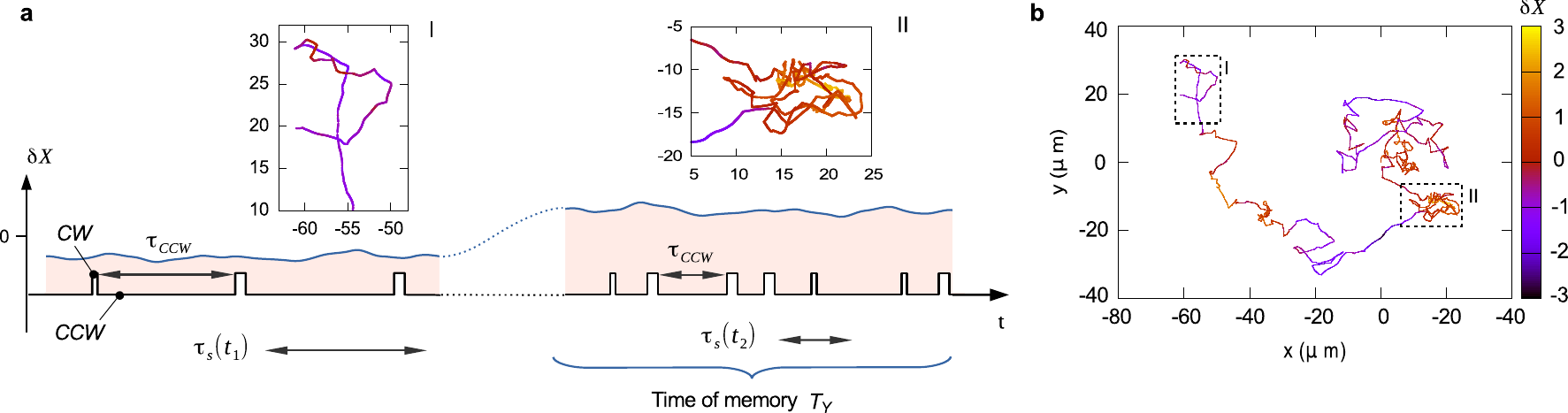}
\caption{Heuristic view of the behavioral variability model.
(a) Time scales of the tumbling process and the CheY-P concentration governing them. The switching time $\tau_\text{s}$ represents the local average of the stochastic run times $\tau_\text{CCW}$. The switching time $\tau_\text{s}$ stays relatively constant during the memory time $T_Y$ and evolves as a function of the normalized CheY-P concentration: $\delta X=([Y]-[Y_0])/\sigma_Y$.
(b) 2D projection of the simulated 3D trajectory where the $\delta X$ fluctuations drive the tumbling process. Insets correspond to different levels of [$\delta X$]: inset I depicts low CheY-P level and inset II depicts high CheY-P level.}
\label{fig:time_scales}
\end{figure*}

To  rationalize and quantify our experimental findings,  we adapt the simple but enlightening physical model proposed by Tu and Grinstein  \cite{Tu2005}. The behavioral variability (BV) model we present here quantifies the role of fluctuations of the phosphorilated protein CheY-P in the regulation of the motor switching statistics. The key idea is that the observed typical switching time at a given moment, depends on the instantaneous CheY-P concentration $[Y](t)$. Then, considering concentration fluctuations around a mean value ($\delta Y(t)= [Y](t)-[Y_0]$), one obtains a two state model with a time varying barrier describing the CCW$\rightarrow$CW switching process. Tu and Grinstein model the $\delta Y$ fluctuations as an Ornstein-Uhlenbeck process with a memory (relaxation) time $T_Y$, hence yielding a Gaussian distribution for $\delta Y$ values. Note that $T_Y$ is considered to be larger than typical motor switching times [see Fig.\ \ref{fig:time_scales}(a) for the relevant time scales].

For small fluctuations of concentration, the average switching time can be written as 

\begin{equation}
\tau_\text{s}=\tau_0 e^{-\Delta_n \delta X}. \label{eq.model}
\end{equation} 
Here, $\delta X$ corresponds to the fluctuations in concentration normalized by the $\delta Y$ standard deviation $\sigma_Y$; $\tau_0$ is a typical switching time corresponding to the mean concentration $[Y_0]$ and $\Delta_n=\alpha \frac{\sigma_Y}{Y_0}$. The parameter $\alpha$  is positive \cite{Cluzel2000} and measures the sensitivity of the switch to variations in $[Y]$. This means that higher concentrations of CheY-P will lead to shorter run times. Note that in principle the two switching times describing CCW$\rightarrow$CW (run times) or  CW$\rightarrow$CCW (tumbling times), could be modeled with corresponding parameters  $\tau_0$ and  $\Delta_n$. However, as the results from Korobkova \textit{et al.} \cite{Korobkova2004} show, in contrast with run times, the distribution of tumble times is exponential, meaning that the equivalent of $\Delta_n$ for tumbles is small. Hence, we will consider the tumbling times as a Poissonian process, well described by a single time scale.

Let us first consider the CCW$\rightarrow$CW switching time distributions. Each observed time belongs to a Poisson distribution with a typical time $\tau_\text{s}$ set by the current CheY-P concentration $[Y](t)$ [see Eq.\ \eqref{eq.model}]. As a consequence, the observed switching statistics for an individual bacterium when observed over a time interval short with respect to the memory time, should approximately appear as an effective Poisson process.  
This is indeed the case, as shown from the collapse of the rescaled orientation correlation functions onto a single exponential decay shown on Fig.~\ref{fig:correlation}(a). 
The model provides a second important outcome. A random choice of a bacterium in a population is like a random choice of  $\delta X$, hence defining a typical switching time $\tau_\text{s}$ for this bacterium. A Gaussian distribution for $\delta X$, as assumed by the behavioral variability (BV) model, leads to a Gaussian distribution of $\ln(\tau_\text{s})$ characterized by an average $\ln(\tau_0)$ and a standard deviation  $\sigma_{\ln\tau_\text{p}}= \Delta_n$, yielding naturally a large log-normal distribution of $\tau_\text{s}$ provided the switch sensitivity $\alpha$ is large. Note that the power law distribution discussed by Tu ann Grinstein \cite{Tu2005} is obtained in the limit of very large $\Delta_n$ and not in contradiction with the above statement.
As $\tau_\text{s}$ and $\tau_\text{p}$ are proportional, the distribution of $\ln(\tau_\text{p})$ should also be Gaussian.

To illustrate this idea, a very long 3D trajectory was synthesized numerically using the switching statistics from the  BV model. Fig.~\ref{fig:time_scales}(b) shows a 2D projection (see Methods section for technical details and Sect.\ \ref{sec.comparisonmodel} for the parameter values). The simulated trajectory contains very persistent (inset I) and very non persistent (inset II) parts. The colors represent the local values of $\delta X$ illustrating the direct influence of the slow variations of CheY-P concentration on the bacterial motility, hence explaining the observed behavioral variability.

\subsection{Memory time}

\begin{figure*}
 \includegraphics[width=\linewidth]{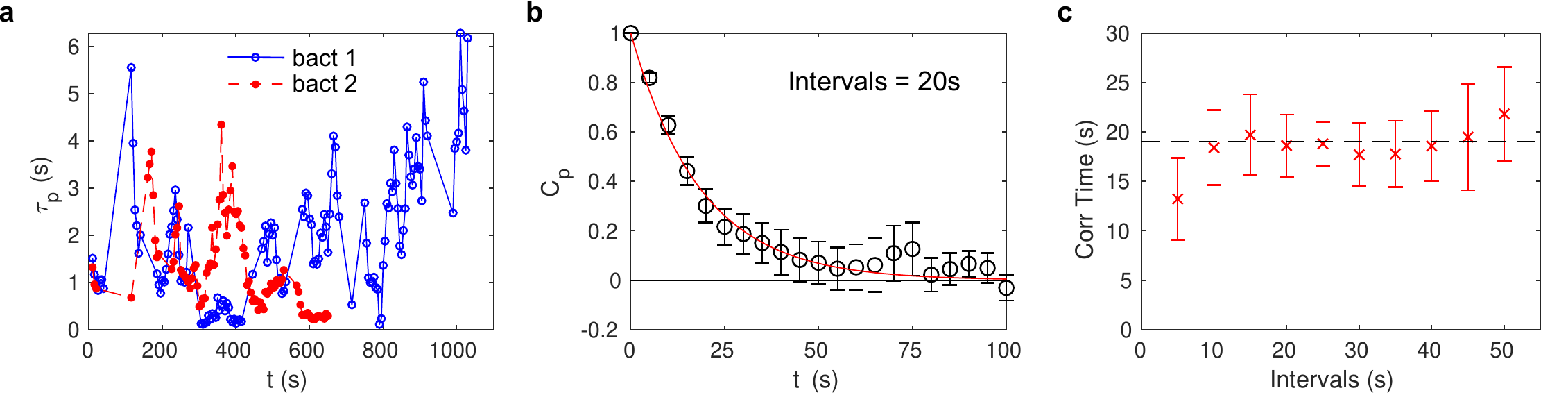}
\caption{Determination of the memory time.
(a) Persistence times $\tau_\text{p}$ computed over pieces of span \SI{20}{\second} and shifted \SI{5}{\second}   for two different bacteria. Gaps larger than $5$\,s between consecutive points correspond to lapses in which the bacteria were swimming close to surfaces. 
(b) Persistence times self-correlation function $C_\text{p}$ using pieces of \SI{20}{\second}. Points represent the average over the ensemble of bacteria.
(c) Correlation time of the persistence times as a function of lengths of the pieces. We extract the memory time to be $T_Y = \SI{19.0 \pm 1.3}{\second}$ . 
}
\label{fig:long_tracks_analysis}
\end{figure*}

The evolution of persistence times $\tau_\text{p}$ along individual trajectories display large variations. It is shown in Fig. \ref{fig:long_tracks_analysis} (a) for the case of two different bacteria continuously tracked for 11 min and 17 min. The values of $ \tau_\text{p} $ for each track were extracted from intervals of span $20$\,s shifted $5$\,s along the trajectory. Gaps larger than $5$\,s between consecutive points correspond to lapses in which the bacterium was swimming close to a surface. Analyzing for example the bacterium of the blue longer trajectory, at time \SI{300}{s} (\SI{5}{min}) it displays a persistence time close to $0.1$\,s, in contrast with a persistence time close to \SI{5}{s} around time  \SI{1000}{s} ($\sim$\SI{17}{\minute}).
This temporal variation of $\tau_\text{p}$ is considered in the framework of the BV model.
The memory time $T_Y$ is then a central parameter of the BV model, as it provides a natural separation between short-time measurements and long time measurements. 
Therefore, for a correct statistical interpretation of the results, $\tau_\text{p}$ values must be extracted from pieces of tracks not longer than the memory time $T_Y$.

We estimate the memory time $T_Y$ from the long time tracking data using the following procedure. For each bacterial trajectory, we compute a sequence of $\tau_\text{p}$ using intervals of specific duration.
For each sequence of $\tau_\text{p}$ we compute the self-correlation function of persistence times, $C_\text{p}(t)=\langle \ln \tau_p(t+t') \ln \tau_\text{p}(t')\rangle$, where the average is done over $t'$. The average of $C_\text{p}$ over the ensemble of trajectories is fitted with an exponential, giving the correlation time [Fig. \ref{fig:long_tracks_analysis} (b)]. 
With this procedure, we investigate different lengths of intervals [Fig.~\ref{fig:long_tracks_analysis}(c)], finding that the correlation times grow with the duration of the interval until saturation at the value of the memory time $T_Y \approx \SI{19.0 \pm 1.3}{\second}$.

\subsection{Comparison with the model} \label{sec.comparisonmodel}

The BV model depends on several parameters: the memory time $T_Y$, the mean switch time and sensitivity $\tau_0$ and $\Delta_n$ respectively, the rotational diffusion coefficient $D_\text{r}^\text{B}$, and the dimensionless rotational diffusion coefficient $\widetilde{D}^\text{eff}_\text{r}$ used to modeling the reorientation during tumble (see Methods for details).
We have determined $T_Y$ from the experiments, while the rest of the parameters are fitted using the following protocol. A long simulated trajectory is generated and cut in pieces of duration \SI{20}{\second}, similar to the analysis of the experimental tracks, and the persistence time is computed for each piece.  We look for the values of the parameters that best agree with the experimental values of the first four moments of the distribution of $\ln\tau_p$. 
The result is
$\tau_0 = \SI{1.53}{\second}$, 
$\Delta_n = 1.62$, 
$D^\text{B}_\text{r} = \SI{0.025}{\second^{-1}}$, 
and $\widetilde{D}^\text{eff}_\text{r} = 3.86$. 
Note that the velocity does not appear in the fit, because we compare simulations and experiments using the persistence times, which depend only on the orientations.

\begin{figure}
	\centering
	\includegraphics[width=0.9\linewidth]{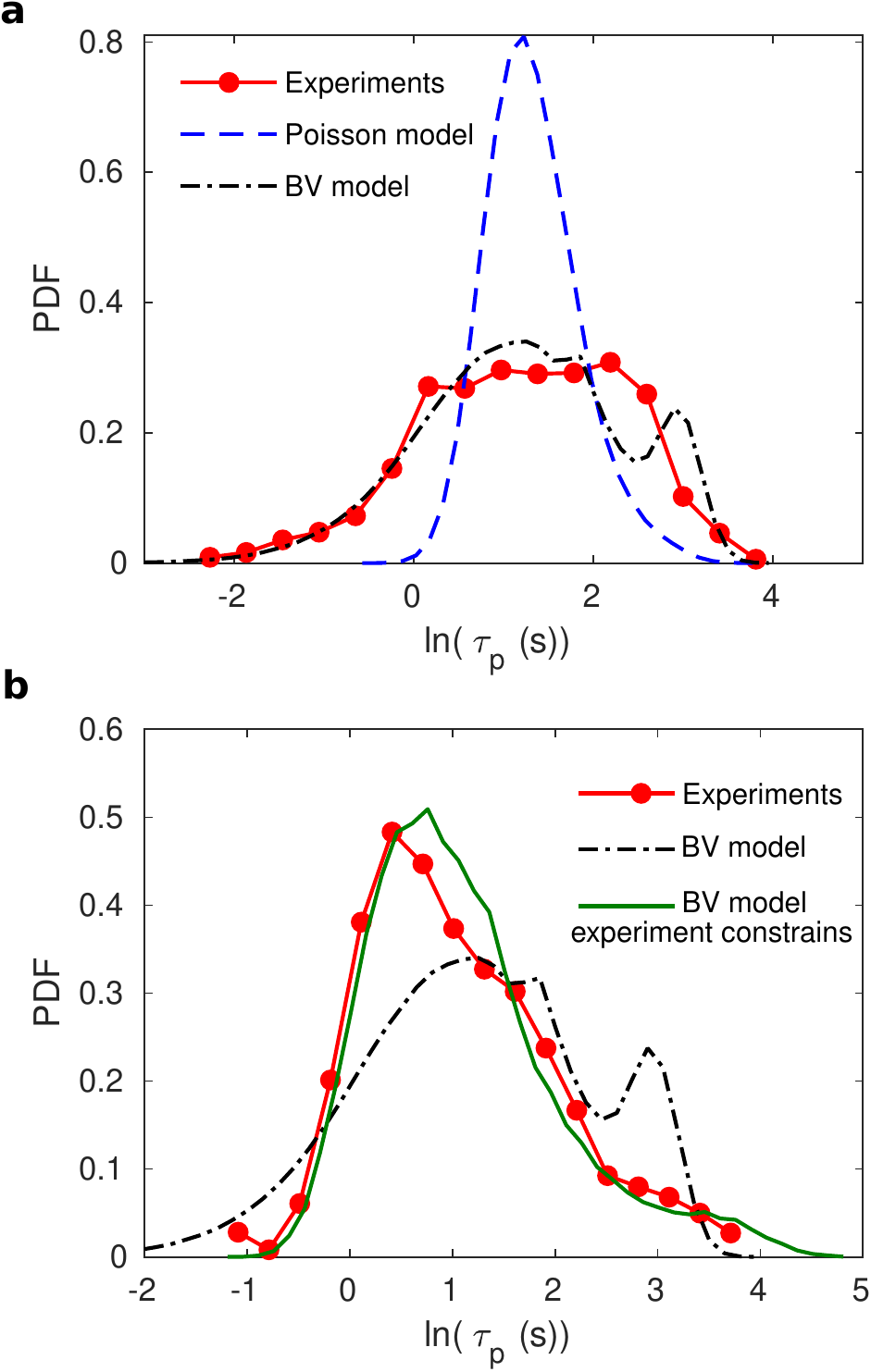}
	\caption{Probability density function of logarithm of persistence times $(\ln \tau_\text{p})$.
			(a) The values $ \tau_{\text{p}} $ were extracted from pieces of track that last 20 s. Simulations using two different models are shown: The Poisson model does not reproduce the experiments, while the behavioral variability (BV) model reproduces the main features. The curve ``BV model'' is the same in both panels. 
			(b) The experimental distribution corresponds to the combined RP437 bacteria in all media, from Fig. \ref{fig:correlation}.
			``BV model experiment constrains'' was determined from the same simulations of  ``BV model", but analyzing pieces of the trajectory that follow the experimental constraints in this case. It reproduces the experiments without additional parameters.}
	\label{fig:PDF}
\end{figure}

Figure \ref{fig:PDF} (a) compares the experimental distribution of $\ln\tau_p$ with the results from simulations using the optimal parameters. The agreement is very good, with two features that need discussion. First, in agreement with the BV model, the distributions are not exactly Gaussian, but present a negative excess kurtosis. 
With 63\% probability, the switch times are in the range $[\tau_0e^{-\Delta_n}, \tau_0e^{\Delta_n}]=[\SI{0.30}{\second},\SI{7.7}{\second}]$.
Hence, there is no complete separation of time scale with $T_Y$. As a consequence, in each piece, $\delta X$ is not constant and the measured and simulation distributions result from the mixture of different values of $\tau_\text{s}$. Note that shorter pieces would imply too few tumble events, and would make it unreliable to use the orientation correlation function. A perfect log-normal distribution could be observed if there was a good separation of time scales, allowing the choice of intervals such that $\tau_0\ll T_\text{interval}\ll T_Y$.

The second feature is the small peak at $\ln\tau_\text{p}\approx 3$ in the simulations. This peak corresponds to pieces of the trajectory where no single tumble took place. The change in orientation is only due to rotational diffusion during a run. Because $\tau_\text{B}=1/2D_r^\text{B}\approx\SI{20}{\second}$ is similar to $T_Y$, no complete reorientation occurs in the interval, resulting in a distribution of $\tau_\text{p}$ for non-tumbling swimmers. 
In fact,  the persistence times for the non-tumbling bacteria [strain CR20, Fig. \ref{fig:correlation}(b)] coincide with this peak. This feature should also be present in experiments, but as discussed in Sect.\ \ref{sec.population}, $D_r^\text{B}$ depends strongly on the bacterial dimensions, which vary within the population. 
 This dispersion of rotational diffusion and other imperfections blur this peak in the experiments, contrary to the simulations, where all swimmers are identical. 
Note that despite of the diversity, the fitted value of $D_r^\text{B}$ matches closely the prediction made in Sect. \ref{sec.population} for ellipsoidal swimmers.

Since the pieces of trajectories are of finite length, the orientation correlation function is not perfectly sampled and, even for a constant switch time $\tau_\text{s}$, the persistence times $\tau_\text{p}$ obtained from an exponential fit would present some dispersion. To test whether the observed dispersion is only due to the data analysis protocol, we perform simulations with a Poisson model. For this, we look for the best parameters to reproduce the first fourth moments of the distribution of $\ln \tau_\text{p}$, setting $\Delta_n=0$. The result is
$\tau_0 = \SI{1.18}{\second}$, 
$D^\text{B}_\text{r} = \SI{0.026}{\second^{-1}}$, 
and $\widetilde{D}^\text{eff}_\text{r} = 1.61$. 
Figure \ref{fig:PDF} (a) presents the resulting distribution, which is far from the experimental one. We conclude that a Poisson process cannot explain the broad distribution of persistence times observed experimentally.

Finally, for consistency, we return to the persistence times obtained in Fig.\ \ref{fig:correlation}. In this experimental protocol the trajectories were selected within a certain height ($10 \mu m$ to $140 \mu m$ from the surface) and longer than $8 s$. 
The corresponding experimental distribution of $ln(\tau_p)$ for RP437 bacteria in all media [Fig. \ref{fig:PDF} (b)] displays a clear positive skewness, which differs strikingly from the experimental measures of panel (a), done using the same bacterial strain and confinement, and similar chemical environment.
This difference originates from a measurement bias  built-in the analysis of panel (b) [and Fig. \ref{fig:correlation}]. The bias is a consequence of a preferential selection of long trajectories staying essentially in the $x$-$y$ plane, with limited bounds in the vertical direction. 
The skewness is enhanced by the broad distribution of run times, since very persistent swimmers will likely quit the measurement region in a very short time, hence privileging small persistence times. 
The black dotted-dashed line is the same in panels (a) and (b), representing the distribution of persistence times from simulations of the BV model that fit the experiments in panel (a). 
When this same simulation is analyzed by taking pieces following strictly the experimental constraints, both on duration and vertical spatial exploration, the resulting distribution (green solid line) compares very well and noticeably without any fitting parameter, to the experimental curve in panel (b).


\section{Conclusions}

We have shown that the 3D spatial exploration of an adapted \textit{E.\,coli} reflects a behavioral variability that we associate with intrinsic noise in the chemotaxis pathway controlling the run-and-tumble sequence. Our results for free swimming bacteria are consistent with models describing motor switching dynamics based on tethered cell measurements. We identified a large log-normal distribution of persistent times stemming from the slow fluctuations of an internal variable accounting for the CheY-P concentration near the motors. In the context of many recent works on statistical physics of active matter, we suggest that this large variability should be included into the description of bacterial fluids. This is expected to influence significantly the computation of averaged quantities like diffusivity, viscosity or any constitutive relations of macroscopic transport processes.

The broad distribution of run times is likely to introduce measurement biases in practical situations. Here, we reduce the bias by taking pieces of trajectories of equal length, not larger than the memory time.
Mixing trajectories of different lengths can result in highly distorted distributions.

The large distribution of motility features is likely to influence the time bacteria spend close to surfaces, 
with consequences for the transport in confined media, where the presence of surfaces is crucial~\cite{Altshuler2013, Figueroa2013, Figueroa2015, figueroa2019coli}. We expect the chemotactic drift to be sensitive to the distribution of CheY-P concentrations, since a non-local spatio-temporal coupling will take place between chemical gradients and bacterial concentration. This should be taken into account in future motility modeling.  Finally, these findings may also impact our vision on how bacterial populations react to environmental changes, colonize space, swarm in a biofilm \cite{Ariel2015} or interact with other communities.

\section{Materials and Methods}

\subsection*{Bacterial strains and culture}
We used the wild type strains RP437 and AB1157 and a smooth swimmer mutant strain CR20 ($\Delta$CheY) expressing YFP (Yellow Fluorescent Protein) from a plasmid. Bacteria were grown overnight at 30$^{\circ}$C in M9G medium [M9 minimal medium supplemented with glucose (\SI{4}{\gram/\liter}), casamino acids (\SI{1}{\gram/\liter}), MgSO$_4$ (\SI{2}{\milli M}) and CaCl$_2$ (\SI{0.1}{\milli M})] plus the corresponding antibiotics, up to optical density = 0.5 at \SI{595}{\nano\meter}. Cells are then washed 3 times by centrifugation at 2000g for \SI{5}{\minute} and suspended in a motility buffer (\SI{10}{\milli M} potassium phosphate buffer pH$\sim$7.0, 0.1mM EDTA, \SI{1}{\micro M} L-methionine and \SI{10}{\milli M} sodium L-lactate), supplemented with polyvinylpyrrolidone (PVP-360kDa 0.002\%) and, when indicated, with L-Serine (\SI{0.04}{\gram/\milli\liter}).

\subsection*{The 3D Lagrangian Tracker}

We developed a device for keeping individual microscopic objects --as swimming bacteria-- in focus, as they move in microfluidic chambers \cite{Darnige2016}.
The system is based on real-time image processing, determining the displacement of the stage to keep the chosen object at a fixed position in the observation frame. The $z$ displacement of the stage is based on the refocusing of the fluorescent object that keeps the moving object in focus. The algorithm for $z$ determination is designed for not being affected by photo-bleaching. 

The instrument is mounted on an epi-fluorescent inverted microscope (Zeiss-Observer, Z1) with a high magnification objective (100 $\times$ /0.9 DIC Zeiss EC Epiplan-Neofluar), a $x$-$y$ mechanically controllable stage with a $z$ piezo-mover
from Applied Scientific Instrumentation (ms-2000-flat-top-xyz) and a digital camera ANDOR iXon 897 EMCCD.
The device works nominally at \SI{30}{fps} on a $512\times512~\si{pix^2}$ matrix but a faster tracking speed of $\SI{80}{\hertz}$ can be achieved reducing the spatial resolution to $128\times128~\si{pix^2}$. It provides images of the object and its track coordinates with respect to the micro-fluidic device.

The tracking limitations come essentially from the $z$ exploration range, restricted by the working distance of $\SI{150}{\micro\meter}$ of the objective. In the $x$-$y$ plane, the spatial limitations are virtually nonexistent, since the stage displacement can be as long as $\SI{15}{\centi\meter}$, which is much bigger than the typical sizes of the sample (a few millimeters).
Details of the apparatus are given in \cite{Darnige2016}, as well as an exhaustive explanation of a method for correcting the mechanical backlash typically affecting these systems and a discussion of the device's performance and limitations.

\subsection*{Experimental geometries and bacteria tracking}

We monitor hundreds of single \textit{E. coli} in a drop of a diluted homogeneous suspension (concentration $\sim \SI{e5}{bact/\milli\liter}$) squeezed between two horizontal glass-slides. The drop has typically a diameter of \SI{1}{\milli\meter}. The gap between the two glass plates is $\SI{250}{\micro\meter}$. 
For the experiments displayed in  Fig.~\ref{fig:correlation}, only pieces of 3D trajectories remaining between the vertical bounds $z_m = \SI{10}{\micro\meter}$ from the bottom surface and $z_M = \SI{140}{\micro\meter}$, the highest possible height and lasting more than \SI{8}{\second} are taken into account. 
For the set of very long tracks of Fig~\ref{fig:shallow_channel}, the gap between the glass plates is also $\SI{250}{\micro\meter}$, but the whole trajectories are captured, as they alternate between bottom and top. For the analysis, only pieces farther than \SI{10}{\micro\meter} from the surfaces are taken into account.

The velocities are determined from second order Savitzky-Golay filtering of the coordinates over \SI{0.1}{\second}, resulting in uncertainties close to $5 \%$ \cite{Figueroa_Thesis}.
For each track, the velocity distribution shows a peak corresponding to the mean run velocity and a low velocity tail corresponding to the contribution of sudden velocity drops (Fig. \ref{fig:velocity_dist}). Peak velocities were in average $\langle V \rangle=\SI{27\pm6}{\micro\meter/\second}$.
To compute the correlation function $C(\Delta t)$, the average is made over time, the lag time is offset by \SI{0.2}{\second} to avoid the short time decorrelation due to wobbling \cite{Figueroa_Thesis, bianchi2017holographic}. The correlation function is then normalized by its value at  \SI{0.2}{\second} to yield 1 at the lag time origin.

\subsection*{Track simulations using the BV model} 

Swimmers are described by their position $\vec r$, orientation $\hat p$, and the instantaneous value of the normalized fluctuations of the CheY-P concentration $\delta X=([Y]-[Y_0])/\sigma_Y$. During run phase, they obey the equation
\begin{align}
\dot{\vec r} &= V \hat{p},\\
\dot{\hat p} &= \sqrt{D_\text{r}^\text{B}} (I-\hat{p} \hat{p}) \vec \eta(t),\\
\dot{\delta X} &= - \delta X/T_Y + \sqrt{2/T_Y}\xi(t),
\end{align}
where $V$ is the swim velocity, $D_\text{r}^\text{B}$ is the rotational diffusion coefficient, $T_Y$ is the memory time,  $(I-\hat{p} \hat{p})$ is a projector orthogonal to $\hat{p}$, $\xi$ is a white noise of zero mean and correlation $\langle \xi(t) \xi(s)\rangle = \delta(t-s)$ and $\vec{\eta}$ is a white noise vector of zero mean, where the components have correlations $\langle \eta_i(t) \eta_k(s)\rangle = \delta_{ik}\delta(t-s)$.

The  BV model yields a relation between the characteristic switching time for the transition CCW$\to$CW (run to tumble) and the CheY-P concentration.
As a simplification, we assume that due to the small cellular dimensions, all 6 flagella operate at the same CheY-P concentration and that the reverse of rotation direction of a single flagellum is enough to trigger a tumble. Hence, the probability to tumble in $\Delta t$ would be $6\Delta t/\tau_\text{s}$. To simplify notation, we absorb the factor 6 into $\tau_0$, resulting in a tumble probability $\Delta te^{\Delta_n \delta X}/\tau_0$.

The BV model predicts that the characteristic switching time for the transition CCW$\to$CW (tumble to run) is also given from an activated process. But, as the corresponding value of $\Delta_n$ is small,  the tumble duration is given by a Poisson process with characteristic time $\tau_1$. In addition, the reorientation dynamics during a tumble needs to be modeled. A priori, the link between motor switch and tumble is far from being trivial as  in principle, one needs to account for the hydrodynamically complex bundling/unbundling process of the multi-flagellated {\it E. coli} bacteria \cite{Darnton2007, mears2014escherichia}.  Here we rather follow a simple effective approach inspired by Saragosti \textit{et al.}~\cite{Saragosti2012}. We model the reorientation dynamics during tumbling as an effective rotational diffusion process with a coefficient $D^\text{eff}_\text{r}$. 
Defining the dimensionless combination $\widetilde{D}^\text{eff}_r=D^\text{eff}_r\tau_1$, the dimensionless tumble durations are sorted from an exponential distribution with a typical time equal to one and, during a tumble, the dynamics is
\begin{align}
\dot{\vec r} &= 0,\\
\dot{\hat p} &= \sqrt{\widetilde{D}^\text{eff}_\text{r}} (I-\hat{p} \hat{p}) \vec \eta(t),\\
\dot{\delta X} &= 0.
\end{align}
After the tumble phase, a new run phase starts.

\acknowledgements
The authors thank Dr Reinaldo Garc\'ia Garc\'ia for useful discussions, Pr Axel Bugin for bacterial strains and Pr Igor Aranson for comments on the manuscript. This work was supported by the ANR grant ``BacFlow'' ANR-15-CE30-0013 and the Franco-Chilean EcosSud Collaborative Program C16E03. N.F.M. thanks the Pierre-Gilles de Gennes Foundation for financial support. A.L. and N.F.M. acknowledge support from the ERC Consolidator Grant PaDyFlow under grant agreement 682367. R.S. acknowledges the Fondecyt grant No.\ 1180791 and Millenium Nucleus Physics of Active Matter of the Millenium Scientific Initiative of the Ministry of Economy, Development and Tourism (Chile). V.A.M was funded by ERC AdG 340877 (PHYSAPS) and Joliot-Curie Chair from ESPCI.

\section*{Appendix A: Persistence correlation function}
The orientation correlation function is defined as
\begin{align}
C(\tau) = \langle \hat{p}(t)\cdot\hat{p}(t+\tau) \rangle =\langle \cos(\theta(\tau)) \rangle,
\end{align}
where $\hat{p}$ is the director vector and the average is done over time $t$.

To compute the correlation function, we use a kinetic theory approach. The object under study is the distribution function $f(\hat{p},t)$, which gives the probability that a bacterium has an orientation $\hat p$ at time $t$. In this context,  the correlation function is obtained assuming that the initial condition at $t=0$ is with the bacterium pointing in a specific direction, say $\hat{p}_0$. Hence, we have to compute $C(\tau) = \langle \hat p(\tau)\cdot \hat p_0\rangle$, where now the average is over the distribution function. At the end, another average, over $\hat p_0$, should be done. In practice  this last average is unnecessary by the isotropy of  space because the first average gives already a value independent of $\hat p_0$.

The distribution function obeys the kinetic equation~\cite{Saintillan2010,Soto}
\begin{align}
\derpar{f}{t} = -L f \label{eq.kineq},
\end{align}
with 
\begin{align}
f(\hat p,0) = \delta(\hat p-\hat p_0) \label{eq.ic}
\end{align}
and $L$ the evolution operator. Two models must be considered. In the case of Brownian rotational diffusivity
\begin{align}
L f = -D_\text{r}^\text{B} \nabla^2_{\hat{p}} f,
\end{align}
where $D_\text{r}^\text{B}$ is the rotational diffusion coefficient and $\nabla^2_{\hat{p}}$ is the angular part of the Laplacian. In the case of tumbling with a characteristic switch time $\tau_{\text{s}}$ 
\begin{align}
Lf = \frac{1}{\tau_{\text{s}}}\left[ f - \int d\hat p' W(\hat  p',\hat p) f(\hat p')\right].
\end{align}
The kernel $W(\hat  p',\hat p)$ gives the probability that for a swimmer with director $\hat p'$, after tumbling, the new director is $\hat p$. It is normalised to $\int d\hat p\, W(\hat  p',\hat p)=1$, indicating that some director $\hat p$ must be chosen.
If the space is isotropic, the kernel only depends on the relative angle between the directors, that is, $W(\hat  p',\hat p)=w(\hat p'\cdot\hat p)$.
Finally, if tumbling and diffusion are present, the operator is just the sum of both.

If the space is isotropic, the evolution operator is also isotropic, which in this case implies that it conmutes with the angular Laplacian, $\nabla^2_{\hat p}$. Therefore, both operators share eigenfunctions, which are the spherical harmonics $Y_{lm}(\hat p)$. Then, there are eigenvalues $\lambda_l$,
\begin{align}
L Y_{lm} = \lambda_l Y_{lm}
\end{align}
that, by isotropy, do not depend on the second index $m$. For the diffusion case, the eigenvalues are known exactly, while for tumbling they are proportional to $1/\tau_{\text{s}}$ and depend on the kernel model. In summary,
\begin{align}
\lambda_l = D_\text{r}^\text{B} l(l+1) +1/(a_l \tau_{\text{s}}),
\end{align}
where $a_l$ are dimensionless parameters of order 1 that depend on the kernel $w$.

Using the basis of the spherical harmonics, the solution of the kinetic equation \reff{eq.kineq}  is
\begin{align}
f(\hat p,t) = \sum_{lm} f_{lm}(0) Y_{lm}(\hat p) e^{-\lambda_l t},
\end{align}
where $f_{lm}(0)$ depend on the initial condition \reff{eq.ic}.
 
Now, the correlation function is
\begin{align}
C(t) &= \langle \hat p(t)\cdot \hat p_0\rangle,\\
&= \int d\hat p\, \hat p_0\cdot \hat p f(\hat p,t),\\
&= \sum_{lm} f_{lm}(0) e^{-\lambda_l t} \hat p_0\cdot \int d\hat p\,  \hat p Y_{lm}(\hat p).
\end{align}
Using that $\hat p$ can be written as a linear combination of $Y_{1m}$, with $m=0,\pm1$ and the orthogonality of the spherical harmonics it is obtained that the integral is not vanishing only for $l=1$. Combining factors, one obtains
\begin{align}
C(t) = C_0 e^{-t/\tau_\text{p}},
\end{align}
where
\begin{align}
\tau_\text{p} = \frac{a_1 \tau_{\text{s}}}{1+a_1 \tau_{\text{s}}/\tau_\text{B}}
\end{align}
and $\tau_\text{B}=1/(2D_\text{r}^\text{B})$ is the Brownian decorrelation time.

In the classical picture, where all bacteria have a single value for $\tau_{\text{s}}$, the decorrelation time $\tau_\text{p}$ is single valued also. When $\tau_{\text{s}}$ is broadly distributed, the decorrelation time $\tau_\text{p}$  will also follow a broad distribution for $\tau_\text{p} \ll \tau_\text{B}$ and it is bounded from above by $\tau_\text{B}$.
Finally, in the description of Berg and Brown \cite{Berg1972}, the tumble angles are distributed with a peak at $63^o$. In this case $a_1=1/(1-\langle\cos\theta\rangle)$ \cite{Saragosti2012}.

\end{document}